\documentclass[letterpaper,titlepage,11pt]{article}

\usepackage{jheppub}

\usepackage[T1]{fontenc}
\usepackage[utf8]{inputenc} % pdfLaTeX için
\usepackage[turkish,english]{babel}

\usepackage{amsmath,amssymb,amsfonts}
\usepackage{esint}
\usepackage{empheq}
\usepackage{multirow}
\usepackage{enumerate}
\usepackage{needspace}

\usepackage{xcolor}
\usepackage{tikz}

\usepackage{textcomp}
\usepackage{bbm}

\usepackage{lineno}
\usepackage[active]{srcltx}

\usepackage{mathrsfs}
\usepackage{booktabs}

\usepackage{caption}

\usepackage{tikz}
\usetikzlibrary{arrows.meta,positioning}
\title{\boldmath{Holonomies and Boundary Symmetries in the Discrete BF Formulation of Carroll Dilaton Gravity
}}
\subheader{ITU Preprint,\,\\ \today}
\newcommand{\itu}{\dagger}

\author[\itu]{H.~T.~\"Ozer}
\emailAdd{ozert@itu.edu.tr}

\author[\itu]{Ayt\"ul~Filiz}
\emailAdd{aytulfiliz@itu.edu.tr}

\affiliation[\itu]{Istanbul Technical University,\,Faculty of Science and Letters,
\,Physics Department,\\34469 Maslak,\,Istanbul,Turkey.}
\abstract{We construct a discrete realization of two--dimensional Carroll dilaton gravity based on a BF--type gauge structure with holonomy variables on lattice links. The bulk theory remains topological, while the physical dynamics is encoded in boundary degrees of freedom. Imposing admissible boundary conditions, we derive the asymptotic symmetry structure directly at the lattice level. The least restrictive conditions yield an affine extension of the Carroll algebra, while additional constraints reduce the symmetry to a conformal sector governed by a discrete Virasoro--type algebra. In the continuum limit, the lattice symmetry structure reproduces the expected affine Carroll algebra together with its conformal reduction. The construction therefore provides the ultra--relativistic counterpart of discrete Jackiw--Teitelboim(JT) gravity.
}
\begin{document}
\maketitle
\flushbottom
%%%%%%%%%%%%%%%%%%%%%%%%%%%%%%%%%%%%%%%%%%%%%%%%%%%%%%%%%%%%%%%%%%%%%%%%%%%%%%%%%%%%%%%%%%%%
%%%%%%%%%%%%%%%%%%%%%%%%%%%%%%%%%%%%%%%%%%%%%%%%%%%%%%%%%%%%%%%%%%%%%%%%%%%%%%%%%%%%%%%%%%%%
%%%%%%%%%%%%        INTRODUCTION        %%%%%%%%%%%%%%%%%%%%%%%%%%%%%%%%%%%%%%%%%%%%%%%%%%%%
%%%%%%%%%%%%%%%%%%%%%%%%%%%%%%%%%%%%%%%%%%%%%%%%%%%%%%%%%%%%%%%%%%%%%%%%%%%%%%%%%%%%%%%%%%%%
%%%%%%%%%%%%%%%%%%%%%%%%%%%%%%%%%%%%%%%%%%%%%%%%%%%%%%%%%%%%%%%%%%%%%%%%%%%%%%%%%%%%%%%%%%%%
\section{Introduction}
\label{sec:Intro}
%1

Two--dimensional dilaton gravity provides a useful laboratory for exploring fundamental aspects of gravitational dynamics, gauge--theoretic descriptions of gravity, and holography. Many models of two--dimensional gravity can be expressed as BF theories or, more generally, as Poisson sigma models, where the bulk dynamics is topological and the physical degrees of freedom reside at the boundary 
\cite{Jackiw1984,Teitelboim1983,Witten1991,Ikeda1994,SchallerStrobl1994,Cattaneo:1999fm}.

%2
The gauge theoretic description of two--dimensional gravity in terms of BF--type theories and Poisson sigma models offers a natural setting for understanding the role of gauge symmetries and boundary structures in gravitational systems \cite{Witten1991,Ikeda1994,SchallerStrobl1994,GrumillerValcarcel2021}. In particular, JT gravity plays a central role due to its connection with near--AdS$_2$ holography and the Schwarzian effective theory \cite{Jackiw1984,Teitelboim1983,AlmheiriPolchinski,MaldacenaStanfordYang}.

%3
Ultra--relativistic limits of gravitational theories and the associated Carrollian geometries have recently attracted significant interest. The Carroll algebra arises from an \.Inönü--Wigner contraction 
of the Poincaré algebra in the limit of vanishing speed of light and defines the symmetry structure underlying Carrollian gravity theories \cite{LevyLeblond1965,Bergshoeff2016,Bagchi2010}.
Two--dimensional Carroll dilaton gravity and its supersymmetric extensions have recently been constructed and studied in detail. In particular, the Carroll--Jackiw--Teitelboim model provides the ultra--relativistic counterpart of JT gravity and admits a BF--type gauge structure involving Carroll gauge fields and dilaton multiplets \cite{GrumillerCarrollSUGRA}.

%4
While the continuum structure of Carroll dilaton gravity is now well understood, its discrete realizations remain largely unexplored. Discrete approaches to gauge and gravity theories offer an alternative perspective on topological structures by describing them in terms of holonomies associated with lattice links and faces. Such discretizations play a central role in lattice gauge theory and related approaches to quantum gravity
\cite{Wilson1974,Kogut1979,Rothe2012}. In particular, discrete descriptions of gravitational systems have been argued to capture structural aspects of diffeomorphism symmetry beyond a simple approximation of continuum
geometries \cite{Dittrich2008}.

%5
In analogy with the discrete description of JT gravity \cite{Ozer:2026qlp}, we construct a lattice realization of two--dimensional Carroll dilaton gravity based on a BF--type gauge structure. 
The theory is expressed in terms of group--valued holonomies associated with the links of a lattice decomposition of the spacetime manifold. While the bulk dynamics remains purely topological, nontrivial structures arise at the boundary.

%6
The use of holonomy variables is particularly natural in topological BF--type theories, since the flatness condition is directly captured by the group structure of lattice holonomies. This makes the holonomy framework well suited for analyzing boundary gauge transformations and the associated symmetry algebras. In this picture, the boundary structure takes the form of a current algebra associated with variations of boundary holonomies.

%7
To our knowledge, a systematic holonomy--based discrete construction of Carroll dilaton gravity together with its affine boundary symmetry and conformal reduction has not been previously established. In this work, we develop a discrete BF--type description of two-dimensional Carroll dilaton gravity and analyze the resulting boundary symmetry structure. In particular, we show that the asymptotic symmetry algebra can be derived directly from boundary holonomies, leading to a discrete Virasoro--type structure without relying on a continuum limit or a non--degenerate invariant bilinear form, where the role of an invariant metric is replaced by a natural pairing between adjoint and coadjoint variables while the corresponding Sugawara construction is consistently realized in the reduced abelian sector.

%8
We analyze admissible boundary conditions and derive the resulting asymptotic symmetry structure directly at the discrete level. The least restrictive boundary conditions lead to an affine extension
of the Carroll algebra, while additional constraints reduce the symmetry structure to a conformal sector governed by a Virasoro--type algebra. The conformal reduction of the affine Carroll current algebra
yields a single boundary current from which a Virasoro--type stress tensor arises through a Sugawara-like construction. In this construction the stress tensor appears as a composite operator built
from the boundary current sector, while the dilaton survives as a conformal singlet of conformal weight zero.

%9
In the continuum limit of the lattice construction, we recover the expected Carroll symmetry structures, showing that the discrete framework yields a consistent realization of Carroll dilaton gravity
and its boundary symmetries.

%10
From this viewpoint, the present construction represents the ultra--relativistic counterpart of the discrete BF description of JT gravity. While relativistic JT gravity is governed by the $\mathfrak{sl}(2,\mathbb{R})$
algebra, the Carroll limit leads to a nilpotent symmetry structure whose affine extension exhibits qualitatively different features in the boundary current algebra.

%11
Thus, the discrete holonomy construction provides a direct bridge between lattice gauge variables and boundary symmetry structures, linking an affine Carroll current algebra to a Virasoro-type conformal symmetry within a fully discrete BF--type framework.

%12
This result indicates that the conformal boundary symmetry of Carroll dilaton gravity arises directly from the lattice current algebra, demonstrating that the discrete BF construction already captures the essential symmetry structure of the continuum theory at the level of boundary holonomies.  

%13
The paper is organized as follows. In Section~\ref{sec:Discrete} we introduce the BF--type description of two--dimensional Carroll dilaton gravity and present its discrete realization in terms of holonomies
associated with a cellular decomposition of the manifold. In Section~\ref{sec:ASatLevel} we analyze the residual gauge transformations that survive at the boundary and derive the corresponding asymptotic symmetry structure at the lattice level, including affine boundary conditions and their conformal reduction to a discrete Virasoro-type algebra. In Section~\ref{sec:discus} we discuss possible extensions
of the discrete Carroll construction and outline directions for future work. Finally, Section~\ref{sec:Conc} summarizes our results and presents our concluding remarks.
Appendix~\ref{app:notation} summarizes the notation and conventions used throughout the
paper, Appendix~\ref{app:summary} collects the boundary data and asymptotic symmetry
structures, and Appendix~\ref{app:discrete_curvature} provides details on the discrete curvature
and the flatness constraint.

%%%%%%%%%%%%%%%%%%%%%%%%%%%%%%%%%%%%%%%%%%%%%%%%%%%%%%%%%%%%%%%%%%%%%%%%%%%%%%%%%%%%%%%%%%
%%%%%%%%%%%%%%%%%%%%%%%%%%%%%%%%%%%%%%%%%%%%%%%%%%%%%%%%%%%%%%%%%%%%%%%%%%%%%%%%%%%%%%%%%%
%%%%%%%%%%           SECTION-II       %%%%%%%%%%%%%%%%%%%%%%%%%%%%%%%%%%%%%%%%%%%%%%%%%%%%
%%%%%%%%%%%%%%%%%%%%%%%%%%%%%%%%%%%%%%%%%%%%%%%%%%%%%%%%%%%%%%%%%%%%%%%%%%%%%%%%%%%%%%%%%%
%%%%%%%%%%%%%%%%%%%%%%%%%%%%%%%%%%%%%%%%%%%%%%%%%%%%%%%%%%%%%%%%%%%%%%%%%%%%%%%%%%%%%%%%%%

\section{Discrete BF formulation of  Carroll Dilaton Gravity }
\label{sec:Discrete}
Carrollian geometries arise as ultra-relativistic limits of relativistic
space-time symmetries and have recently attracted considerable attention
in the context of flat space holography and non-Lorentzian gravitational
systems \cite{LevyLeblond1965,Bergshoeff2016,Bagchi2010}.
In two space-time dimensions, Carroll gravity exhibits a particularly
simple structure and can be described in terms of a gauge theory closely
related to BF--type models \cite{GrumillerCarrollSUGRA}.  This perspective is especially
useful for dilaton gravity, where the absence of local propagating degrees
of freedom allows the dynamics to be captured entirely by gauge and
topological data.

In this section we review the bosonic Carroll dilaton gravity and describe
how its gauge structure naturally leads to a holonomy-based description.
This viewpoint provides a direct route toward a discrete realization of
the theory analogous to the discrete description of two--dimensional
JT gravity. For clarity and to avoid notational ambiguity, we summarize all conventions 
and symbols in Appendix~\ref{app:notation}.

The underlying symmetry algebra is the three--dimensional Carroll algebra
generated by a temporal translation $\tt H$, a spatial translation $\tt P$, and
a Carroll boost $\tt B$. The non--vanishing commutation relation is

\begin{equation}
\tt [B,P] = H ,
\end{equation}

while the remaining brackets vanish,

\begin{equation}
\tt [B,H] = 0, \qquad [P,H] = 0 .
\end{equation}

This algebra can be obtained as an ultra--relativistic
\.Inönü--Wigner contraction of the three--dimensional
Poincaré algebra $\mathfrak{iso}(1,1)$ \cite{Inonu:1953sp,LevyLeblond1965}.
Modern developments and applications of Carroll symmetries
in gravitational theories can be found in
\cite{Bagchi2010,Bergshoeff2016}.

The Poincaré algebra  $\mathfrak{iso}(1,1)$ is generated by time translations $\tt H$,
spatial translations $\tt P$, and Lorentz boosts $\tt J$, with
commutation relations
\begin{equation}
\tt [J,P] = H, \qquad [J,H] = P, \qquad [P,H] = 0 .
\end{equation}
Introducing the rescaling
\begin{equation}
\tt J \rightarrow B, \qquad
\tt P \rightarrow P, \qquad
\tt H \rightarrow cH ,
\end{equation}
and taking the limit $c\rightarrow 0$, the commutation relations reduce to
\begin{equation}
\tt [B,P] = H, \qquad
\tt [B,H] = 0, \qquad
\tt [P,H] = 0 ,
\end{equation}
which define the three--dimensional Carroll algebra used in this work. An important structural feature of the resulting algebra is
its nilpotent character, which simplifies the structure of the associated gauge theory and will play an important role in the
discrete construction discussed below. Following the standard gauge-theoretic approach, we introduce a Lie--algebra valued connection one-form
\begin{equation}
\mathcal A = \tau \tt H + e\tt P + \omega\tt B ,
\end{equation}
where $\tau$ plays the role of a temporal vielbein, $e$ is the spatial vielbein, and $\omega$ denotes the Carroll boost connection. The
associated field strength is given by
\begin{equation}
\mathcal F = d \mathcal A +\mathcal A \wedge\mathcal A .
\end{equation}
Using the Carroll commutation relations one obtains the curvature
components
\begin{align}
\mathcal F_H &= d\tau + \omega \wedge e , \\
\mathcal F_P &= de , \\
\mathcal F_B &= d\omega .
\end{align}
As in other dilaton gravity models \cite{Jackiw1984,Teitelboim1983}, the theory is completed by introducing a Lie-algebra valued dilaton
multiplet
\begin{equation}
\mathcal X =\mathcal  X_{\tt H}  \tt H +\mathcal  X_{\tt P} \tt P +\mathcal X_{\tt B}  \tt B .
\end{equation}
%\paragraph{Invariant pairing}
\newpage
A natural starting point for two-dimensional gravity is its realization
either in a Chern--Simons-like framework or, more appropriately in two dimensions,
as a BF theory underlying JT gravity. In the Chern--Simons framework, the existence
of a non-degenerate invariant bilinear form on the gauge algebra is essential for
defining the action. While this requirement is satisfied for relativistic algebras
such as $\mathfrak{sl}(2,\mathbb{R})$, it becomes problematic for non-semisimple
algebras, including those arising in Carrollian limits, where invariant metrics are
typically degenerate or do not exist.

In contrast, BF theories do not require a non-degenerate invariant metric. Instead,
the action is defined through the natural pairing between adjoint and coadjoint
variables: the dilaton field $\mathcal X$ takes values in the coadjoint
representation, while the curvature $\mathcal F$ is adjoint-valued, and the action
is constructed from their canonical pairing. This makes the BF framework
particularly well suited for Carroll-type gravitational systems.

Accordingly, we do not assume the existence of an invariant bilinear form on the
algebra. Rather, we work with a pairing compatible with the adjoint--coadjoint
structure, which is sufficient to ensure gauge invariance. Under an infinitesimal
gauge transformation with parameter $\lambda$, one has
\begin{equation}
\delta_\lambda \mathcal X = [\mathcal X,\lambda],
\qquad
\delta_\lambda \mathcal F = [\mathcal F,\lambda],
\end{equation}
so that
\begin{equation}
\label{cancel}
\delta_\lambda \langle \mathcal X,\mathcal F \rangle
\langle [\mathcal X,\lambda],\mathcal F\rangle
+
\langle \mathcal X,[\mathcal F,\lambda]\rangle
=0,
\end{equation}
where the cancellation follows from the dual pairing between adjoint and coadjoint
representations. No invariant metric in the usual Lie-algebraic sense is required.

For non-semisimple algebras such as the Carroll algebra, the Killing form is
degenerate and cannot be used to define the action. Nevertheless, the BF
construction only requires a pairing compatible with the adjoint--coadjoint
structure, which need not coincide with the Killing form nor satisfy
non-degeneracy conditions on the algebra itself.

For practical purposes, we introduce a simple diagonal pairing on the chosen basis,
\begin{equation}
\langle \tt H,\tt H\rangle=\eta^{\tt HH},\qquad
\langle \tt P,\tt P\rangle=\eta^{\tt PP},\qquad
\langle \tt B,\tt B\rangle=\eta^{\tt BB},
\end{equation}
with all mixed components vanishing. The coefficients $\eta^{IJ}$ are fixed
constants specifying the pairing and should be viewed as defining data of the BF
theory, rather than components of an invariant metric on the Carroll algebra. In
particular, allowing $\langle\tt H,\tt H\rangle\neq 0$ is consistent within this
framework.

This structure provides a consistent and sufficient basis for defining the BF-type
action, constructing boundary charges, and analyzing asymptotic symmetries. The
key point is that gauge invariance is ensured by the adjoint--coadjoint pairing,
rather than by the existence of an invariant metric.

The corresponding BF--type action is given by
\begin{equation}
\mathcal S = \int \langle\mathcal X ,\mathcal F \rangle.
\end{equation}
In terms of components the action becomes
\begin{equation}
\mathcal S =
\int
\left(
\mathcal  X_{\tt H} (d\tau + \omega \wedge e)
+
\mathcal  X_{\tt P}\, de
+
\mathcal  X_{\tt B}\, d\omega
\right).
\end{equation}

The equations of motion enforce flatness of the connection together
with covariant constancy of the dilaton multiplet. As a consequence,
the bulk theory possesses no local propagating degrees of freedom and
remains purely topological \cite{Witten1991,Ikeda1994,SchallerStrobl1994}.
The topological nature of the BF--type description makes Carroll
dilaton gravity particularly well suited for a holonomy--based
construction. The gauge field can locally be written in terms of group
elements, so that the physical information becomes naturally carried
by holonomies associated with curves on the manifold. In the Carroll
case this description is further simplified by the nilpotent structure
of the Carroll algebra, which leads to a truncated
Baker--Campbell--Hausdorff expansion and allows holonomies to be
evaluated explicitly. This feature provides a natural starting point
for the discrete construction developed below, where the nontrivial 
physical information resides in boundary data in close analogy
with BF descriptions of two-dimensional dilaton gravity.

This structure makes Carroll dilaton gravity particularly well suited
for a discrete description. 
Rather than viewing discretization as an approximation of a continuum
geometry, it is natural in topological gauge theories to interpret it
as a structural implementation of the underlying gauge symmetries at
the lattice level \cite{Dittrich2008}.
In this perspective,
the fundamental variables are holonomies associated with links of a
cellular decomposition, while the flatness constraints of the BF
theory can be imposed exactly.

Let $\mathcal{M}$ be an oriented two--dimensional manifold with
boundary, whose topology we take to be that of a disk. We discretize
$\mathcal{M}$ by means of a cellular decomposition into vertices
$v$, oriented links $\ell$, and plaquettes $f$. No metric structure
is assumed on the lattice, reflecting the topological nature of the
underlying gauge theory.

In the discrete gauge description the fundamental variables are
group--valued holonomies associated with oriented links $\ell$ \cite{Wilson1974,Kogut1979,Rothe2012},
\begin{equation}
\mathcal U_\ell =\mathcal P \exp\!\left(\int_\ell\mathcal A \right),
\end{equation}
which replace the continuum gauge field. Orientation reversal of a
link corresponds to group inversion,
\begin{equation}
\mathcal U_{\ell^{-1}} = \mathcal U_\ell^{-1}.
\end{equation}

Discrete curvature is represented by the plaquette holonomy
\begin{equation}
\mathcal W_f = \prod_{\ell \in \partial f}\mathcal U_\ell ,
\end{equation}
where the ordered product follows the orientation of the plaquette
boundary. In the continuum limit this expression reduces to the
exponential of the curvature integrated over the plaquette, so that
the condition $\mathcal W_f = 1$ corresponds to vanishing curvature.

In analogy with the continuum BF description, the dilaton field
acts as a Lagrange multiplier enforcing flatness of the connection.
In the discrete theory we therefore associate a Lie--algebra valued
variable $\mathcal X_f$ to each plaquette.

The resulting discrete BF--type action takes the form
\begin{equation}
\mathcal S_{\text{disc}}
=
\sum_f
\left\langle
\mathcal X_f ,
\log\mathcal W_f
\right\rangle
+\mathcal S_{\text{boundary}},
\end{equation}
where $\log\mathcal W_f$ maps the plaquette holonomy to the Lie algebra.

Strictly speaking, the logarithm is well-defined only in a neighborhood of the identity element of the gauge group. In the present construction, this restriction is consistent with the classical solution space of the BF theory, where the flatness condition implies that all plaquette holonomies are continuously connected to the identity. Therefore, the action should be understood as defined within this sector of the configuration space.

%\paragraph{Holonomy logarithm}
The logarithm appearing in the discrete action is defined in a neighbourhood of the identity element of the gauge group. 
Since the configurations considered here correspond to flat connections in the bulk, the plaquette holonomies remain continuously connected to the identity. 
This ensures that the branch of the logarithm used in the action is well defined.
Variation with respect to $\mathcal X_f$ enforces the flatness constraint
\begin{equation}
\mathcal W_f = \mathbbm{1} ,
\end{equation}
for all plaquettes $f$. Together with the lattice implementation of covariant constancy, these conditions eliminate local bulk degrees of
freedom and ensure that the theory remains topological in the bulk, while the nontrivial phase space resides at the boundary.

Finally, the nilpotent structure of the Carroll algebra leads to a significant simplification of the Baker--Campbell--Hausdorff expansion
entering the logarithm of the plaquette holonomy. Nested commutators truncate after finitely many terms, making the explicit evaluation of
holonomies considerably simpler than in relativistic cases. This property will play an important role in the analysis of boundary
symmetries and their discrete realization developed in the following sections. The relation between plaquette holonomies and the continuum 
curvature, together with the derivation of the discrete flatness condition, is presented in Appendix~\ref{app:discrete_curvature}.

To our knowledge, such a holonomy-based discrete description has not been systematically explored for Carroll dilaton gravity.
The construction developed here therefore provides a natural extension of the discrete BF approach to the Carrollian
regime and establishes a direct bridge between Carroll dilaton gravity and lattice gauge descriptions.
 In the following section we analyze the residual gauge transformations that survive at the boundary and give rise to the asymptotic symmetry structure 
of the discrete theory.
%%%%%%%%%%%%%%%%%%%%%%%%%%%%%%%%%%%%%%%%%%%%%%%%%%%%%%%%%%%%%%%%%%%%%%%%%%%%%%%%%%%%%%%%%%
%%%%%%%%%%%%%%%%%%%%%%%%%%%%%%%%%%%%%%%%%%%%%%%%%%%%%%%%%%%%%%%%%%%%%%%%%%%%%%%%%%%%%%%%%%
%%%%%%%%%%           SECTION-II-1     %%%%%%%%%%%%%%%%%%%%%%%%%%%%%%%%%%%%%%%%%%%%%%%%%%%%
%%%%%%%%%%%%%%%%%%%%%%%%%%%%%%%%%%%%%%%%%%%%%%%%%%%%%%%%%%%%%%%%%%%%%%%%%%%%%%%%%%%%%%%%%%
%%%%%%%%%%%%%%%%%%%%%%%%%%%%%%%%%%%%%%%%%%%%%%%%%%%%%%%%%%%%%%%%%%%%%%%%%%%%%%%%%%%%%%%%%%
\subsection{Geometric interpretation of the lattice variables}
\label{sec:Geometric}
The discrete BF--type description introduced above admits a simple geometric interpretation in terms of the cellular decomposition of
the two--dimensional manifold.

Vertices of the lattice represent local gauge frames, while oriented links carry group-valued holonomies encoding the parallel transport
of Carroll gauge fields between neighboring sites. The plaquettes of the lattice capture the discrete curvature through the associated
face holonomies.

From this perspective the discrete variables have a direct geometric meaning: the holonomies represent the transport of Carroll frames along
the lattice links, while the dilaton variables associated with plaquettes enforce the flatness constraint of the BF--type structure.

Because the underlying Carroll algebra is nilpotent, the Baker--Campbell--Hausdorff expansion truncates after finitely many terms.
This property considerably simplifies the evaluation of holonomies compared with relativistic BF theories and plays an important
role in the discrete construction developed in this work.
%%%%%%%%%%%%%%%%%%%%%%%%%%%%%%%%%%%%%%%%%%%%%%%%%%%%%%%%%%%%%%%%%%%%%%%%%%%%%%%%%%%%%%%%%%
%%%%%%%%%%%%%%%%%%%%%%%%%%%%%%%%%%%%%%%%%%%%%%%%%%%%%%%%%%%%%%%%%%%%%%%%%%%%%%%%%%%%%%%%%%
%%%%%%%%%%           SECTION-III     %%%%%%%%%%%%%%%%%%%%%%%%%%%%%%%%%%%%%%%%%%%%%%%%%%%%
%%%%%%%%%%%%%%%%%%%%%%%%%%%%%%%%%%%%%%%%%%%%%%%%%%%%%%%%%%%%%%%%%%%%%%%%%%%%%%%%%%%%%%%%%%
%%%%%%%%%%%%%%%%%%%%%%%%%%%%%%%%%%%%%%%%%%%%%%%%%%%%%%%%%%%%%%%%%%%%%%%%%%%%%%%%%%%%%%%%%%
\section{Asymptotic symmetries at the lattice level}
\label{sec:ASatLevel}
In the discrete BF--type description introduced in the previous section, asymptotic symmetries arise as residual gauge
transformations that preserve a chosen set of boundary conditions, in close analogy with the continuum analysis of
gravitational boundary symmetries \cite{BrownHenneaux1986,GrumillerValcarcel2021}. A key difference with respect to the
continuum analysis of dilaton gravity is that, in the discrete setting, this mechanism is realized directly in terms of
lattice variables and boundary holonomies, without reference to fall-off conditions or continuum gauge fixing.

The bulk theory remains topological due to the flatness constraints imposed by the discrete BF action. 
This follows directly from the flatness constraints of the BF theory, which eliminate all local bulk degrees of freedom.
As a result, bulk gauge transformations act trivially on physical states, while gauge transformations that act nontrivially on boundary links survive
as genuine symmetries of the theory. Consequently, all nontrivial dynamical information becomes localized in boundary holonomies and the associated boundary fields.

In this section we analyze the residual gauge transformations that preserve the boundary conditions and determine the resulting symmetry
structure at the boundary of the lattice. These transformations act on the discrete boundary variables in a manner analogous to the
continuum treatment of asymptotic symmetries in two--dimensional dilaton gravity, but the entire construction is implemented directly
at the lattice level.

The boundary of the discretized manifold is represented by an oriented sequence of lattice links forming a closed chain.
Boundary holonomies associated with these links define the physical phase space of the theory. Residual gauge transformations act on
these holonomies as local symmetry transformations along the boundary lattice.

Depending on the choice of boundary conditions, different symmetry structures can emerge. The least restrictive choice leads to affine
boundary symmetries, corresponding to a lattice realization of an affine current algebra associated with the Carroll algebra.
More restrictive boundary conditions reduce this structure to a conformal symmetry, producing a discrete version of a Virasoro-type
algebra governing the boundary dynamics.

The analysis proceeds in two steps. We first consider affine boundary conditions and derive the associated boundary symmetry algebra.
We then impose additional constraints that reduce the affine structure to a conformal one, leading to the emergence of
discrete Virasoro generators.

%%%%%%%%%%%%%%%%%%%%%%%%%%%%%%%%%%%%%%%%%%%%%%%%%%%%%%%%%%%%%%%%%%%%%%%%%%%%%%%%%%%%%%%%%%
%%%%%%%%%%%%%%%%%%%%%%%%%%%%%%%%%%%%%%%%%%%%%%%%%%%%%%%%%%%%%%%%%%%%%%%%%%%%%%%%%%%%%%%%%%
%%%%%%%%%%           SECTION-III-1     %%%%%%%%%%%%%%%%%%%%%%%%%%%%%%%%%%%%%%%%%%%%%%%%%%%%
%%%%%%%%%%%%%%%%%%%%%%%%%%%%%%%%%%%%%%%%%%%%%%%%%%%%%%%%%%%%%%%%%%%%%%%%%%%%%%%%%%%%%%%%%%
%%%%%%%%%%%%%%%%%%%%%%%%%%%%%%%%%%%%%%%%%%%%%%%%%%%%%%%%%%%%%%%%%%%%%%%%%%%%%%%%%%%%%%%%%%
\subsection{Affine boundary conditions}
\label{sec:Affinebc}
We begin by imposing affine boundary conditions on the discrete boundary holonomies. These conditions represent the least restrictive
choice compatible with a well-defined variational principle while preserving the full residual gauge symmetry at the boundary.
At the lattice level, bulk gauge transformations are rendered pure gauge by the flatness constraints of the BF--type theory, whereas
transformations acting nontrivially on boundary links survive as physical symmetries. Consequently, the nontrivial dynamical content
of the theory becomes localized on boundary holonomies and the associated boundary fields.

In analogy with the standard radial gauge used in asymptotically
AdS$_2$ spacetimes, the Carroll BF connection can be written as

\begin{equation}
\mathcal A = b^{-1} a(\tau) b + b^{-1} d b,
\qquad
\mathcal X = b^{-1} x(\tau) b ,
\end{equation}

where $b(\rho)$ is a group element depending only on the radial
coordinate and independent of the dynamical state.
For definiteness one may choose

\begin{equation}
b(\rho)=e^{\rho \tt B},
\end{equation}

with $B$ the Carroll boost generator.
In this parametrization the reduced connection
$a(\tau)$ takes values in the Carroll algebra and is
independent of the radial coordinate,

\begin{equation}
a(\tau)=a_\tau(\tau)d\tau .
\end{equation}

The dynamical information of the theory is therefore encoded
entirely in the boundary fields contained in $a(\tau)$ and
in the dilaton multiplet $x(\tau)$.

The boundary of the discretized manifold is represented by an oriented circle parametrized by lattice sites 
$n=0,\dots,N-1$, 
with lattice spacing
$
\Delta\tau=\frac{2\pi}{N}.
$
Boundary links connect neighboring sites $(n,n+1)$ and are associated with holonomies
\begin{equation}
\mathcal U_n=\mathcal P\exp\!\left(\int_{\tau_n}^{\tau_{n+1}} a_\tau(\tau)d\tau\right),
\qquad
\tau_n=n\Delta\tau .
\end{equation}

For sufficiently small lattice spacing the boundary connection may be approximated as constant along each link, so that
\begin{equation}
\int_{\tau_n}^{\tau_{n+1}} a_\tau(\tau)\, d\tau
\simeq
\Delta\tau\, a_\tau(\tau_n).
\end{equation}
This motivates the introduction of a logarithmic link variable defined by
\begin{equation}
U_n=\exp(\Delta\tau a_n),
\qquad
a_n = \frac{1}{\Delta\tau}\log U_n \in { \text{Carroll algebra}}
\end{equation}

In the continuum limit one then has
\begin{equation}
a_n = a_\tau(\tau_n)+\mathcal O(\Delta\tau).
\end{equation}

The affine boundary condition is implemented by restricting the boundary connection to the form
\begin{equation}
a_n=\alpha^{_I} \mathcal L^{_I}_n {\tt T}_{_I} ,
\qquad I\in\{\tt H,\tt P,\tt B\},
\end{equation}
where $\tt T_{_{\it I}}$ are generators of the Carroll algebra, $\alpha^{_{\it I}}$  are fixed constants specifying the boundary condition, and $\mathcal L^{_I}_n$  denote the
boundary fields defined on the lattice at $\tau_n=n\Delta\tau$.

Gauge transformations act on the boundary holonomies according to
\begin{equation}
U_n\rightarrow g_n^{-1}U_n g_{n+1},
\qquad g_n\in G_{{ \text{Carroll}}} .
\end{equation}
For infinitesimal transformations
\begin{equation}
g_n=\exp(\lambda_n),
\qquad
\lambda_n\in{\text{Carroll algebra}},
\end{equation}
the transformation becomes
\begin{equation}
U_n\rightarrow e^{-\lambda_n}e^{\Delta\tau a_n}e^{\lambda_{n+1}} .
\end{equation}

Using the Baker--Campbell--Hausdorff expansion to first order gives
\begin{equation}
e^{-\lambda_n}e^{\Delta\tau a_n}e^{\lambda_{n+1}}
=
\exp\!\left(
\Delta\tau a_n
+
(\lambda_{n+1}-\lambda_n)
+
\Delta\tau[a_n,\lambda_n]
+
\mathcal{O}(\lambda^2)
\right).
\end{equation}

Taking the logarithm yields
\begin{equation}
\delta a_n=
\frac{\lambda_{n+1}-\lambda_n}{\Delta\tau}
+
[a_n,\lambda_n].
\end{equation}

This identifies the gauge-covariant discrete derivative
\begin{equation}
\nabla\lambda_n=
\frac{\lambda_{n+1}-\lambda_n}{\Delta\tau},
\end{equation}
so that the transformation law becomes
\begin{equation}
\delta a_n=\nabla\lambda_n+[a_n,\lambda_n].
\end{equation}

These transformations represent the discrete counterpart of continuum affine current transformations and preserve the affine boundary condition provided the gauge parameter takes the form
\begin{equation}
\lambda_n=\epsilon^{_{\it I}}_n \tt T_{_{\it I}} ,
\end{equation}
with arbitrary functions $\epsilon^{_{\it I}}_n$ defined on the boundary lattice.

Substituting the affine boundary condition into the transformation law yields the transformation rules 
\begin{equation}
\alpha^{_{\it I}}\,\delta\mathcal L_n^{_{\it I}}
=
\nabla\epsilon_n^{_{\it I}}
+
f^{{_{\it I}}}{}_{{_{\it J}}{_{\it K}}}\,\alpha^{_{\it J}}\mathcal L_n^{_{\it J}}\epsilon_n^{_{\it K}}.
\end{equation}
Using the only non-vanishing Carroll structure constants
\begin{equation}
f^{\tt H}{}_{\tt B \tt P}=1,
\qquad
f^{\tt H}{}_{\tt P \tt B}=-1,
\end{equation}
the component transformations become
\begin{align}
\delta\mathcal L_n^{\tt H}
&=
\frac{1}{\alpha^{\tt H}}\nabla\epsilon_n^{\tt H}
+
\frac{\alpha^{\tt B}}{\alpha^{\tt H}}\mathcal L_n^{\tt B}\epsilon_n^{\tt P}
-
\frac{\alpha^{\tt P}}{\alpha^{\tt H}}\mathcal L_n^{\tt P}\epsilon_n^{\tt B},\\
\delta \mathcal L_n^{\tt P}
&=
\frac{1}{\alpha^{\tt P}}\nabla\epsilon_n^{\tt P},\\
\delta\mathcal L_n^{\tt B}
&=
\frac{1}{\alpha^{\tt B}}\nabla\epsilon_n^{\tt B}.
\end{align}
Thus only the $\tt H$ -component mixes nontrivially with the remaining boundary fields, while the $\tt P$ and $\tt B$ -components behave as
abelian currents up to derivative terms.
%%%%%%%%Section 3.1 devamı — boundary charges ve affine cebir

The generators of these boundary transformations are determined by the discrete BF symplectic structure. In complete analogy with the lattice
BF description, we require that the charge $Q[\epsilon]$ generate the variation of the boundary connection through the Poisson bracket.
Its variation is therefore taken to be
\begin{equation}
\delta Q[\epsilon]
=
\frac{\tt k}{2\pi}
\sum_n
\Delta\tau\,
\langle
\lambda_n,\delta a_n
\rangle ,
\end{equation}
where $\langle\cdot,\cdot\rangle$ denotes the bilinear pairing on the Carroll algebra.

Assuming integrability, this yields the boundary charge
\begin{equation}
Q[\epsilon]
=
\frac{\tt k}{2\pi}
\sum_n
\Delta\tau
\left(
\alpha^{\tt H} \epsilon^{\tt H}_n \mathcal L^{\tt H}_n
+
\alpha^{\tt P} \epsilon^{\tt P}_n\mathcal L^{\tt P}_n
+
\alpha^{\tt B} \epsilon^{\tt B}_n\mathcal L^{\tt B}_n
\right).
\end{equation}

By definition, the charge generates the affine boundary transformations according to
\begin{equation}
\delta_{\epsilon} a_n
=
\{a_n,Q[\epsilon]\}.
\end{equation}
Comparing this relation with the component variations derived above determines the Poisson brackets of the boundary fields.
\color{black}

Using the Carroll structure constants, the non-vanishing Poisson brackets take the form
\begin{align}
\frac{\Delta\tau}{2\pi}\{\mathcal L^{\tt H}_n,\mathcal L^{\tt P}_m\} &= 0,\\
\frac{\Delta\tau}{2\pi}\{\mathcal L^{\tt H}_n,\mathcal L^{\tt B}_m\} &= 0,\\
\frac{\Delta\tau}{2\pi}\{\mathcal L^{\tt P}_n,\mathcal L^{\tt B}_m\} &= \mathcal L^{\tt H}_n\,\delta_{n,m},
\end{align}
together with the derivative contributions generated by the affine extension,
\begin{align}
\frac{\Delta\tau}{2\pi}\{\mathcal L^{\tt H}_n,\mathcal L^{\tt H}_m\} &= \frac{\tt k}{2}\,\eta^{{\tt H}{\tt H}}\,\nabla\delta_{n,m},\\
\frac{\Delta\tau}{2\pi}\{\mathcal L^{\tt P}_n,\mathcal L^{\tt P}_m\} &= \frac{\tt k}{2}\,\eta^{{\tt P}{\tt P}}\,\nabla\delta_{n,m},\\
\frac{\Delta\tau}{2\pi}\{\mathcal L^{\tt B}_n,\mathcal L^{\tt B}_m\} &= \frac{\tt k}{2}\,\eta^{{\tt B}{\tt B}}\,\nabla\delta_{n,m},
\end{align}
and
\begin{equation}
\frac{\Delta\tau}{2\pi}\{\mathcal L^I_n,\mathcal L^J_m\}
=
f^{IJ}{}_{K}\mathcal L^K_n\,\delta_{n,m}
+
\frac{k}{2}\eta^{IJ}\nabla\delta_{n,m}.
\end{equation}
Here $f^{IJ}{}_{K}$ are the structure constants of the Carroll algebra,
with the only non-vanishing components
\begin{equation}
f^{\tt BP}{}_{\tt H}=1,
\qquad
f^{\tt PB}{}_{\tt H}=-1 .
\end{equation}
For the remainder of the analysis, we adopt the canonical normalization
\begin{equation}
\alpha^{\tt H}=\alpha^{\tt P}=\alpha^{\tt B}=1 .
\end{equation}
With this choice, the boundary currents inherit the standard normalization
of the Carroll structure constants, and no additional normalization scale
is introduced in the affine current algebra. Accordingly, the site-basis
algebra above is understood in this canonical normalization.

These relations define the discrete affine extension of the Carroll algebra
at the boundary. In contrast to the relativistic affine
$\mathfrak{sl}(2,\mathbb{R})$ case, the present algebra inherits the
nilpotent structure of the Carroll algebra itself. Accordingly, the only
non-trivial Lie-algebraic mixing occurs in the $\tt H$-sector, while the $\tt P$
and $\tt B$ components behave as abelian currents up to the central derivative
term.

\paragraph{Discrete Fourier representation.}
The algebra in Eq.~(3.31) is written in the site basis, where
$n,m=0,\ldots,N-1$ label the sites of the periodic boundary lattice.
To make the affine mode structure explicit, we introduce the discrete
Fourier expansion
\begin{equation}
\mathcal{L}^{I}_{n}
=
\sum_{r=0}^{N-1}
\mathcal{J}^{I}_{r}\,e^{ir\tau_n},
\qquad
\mathcal{J}^{I}_{r}
=
\frac{1}{N}
\sum_{n=0}^{N-1}
\mathcal{L}^{I}_{n}\,e^{-ir\tau_n},
\label{eq:CarrollDiscreteFourier}
\end{equation}
Here $I,J\in\{\tt H,\tt P,\tt B\}$ remain the internal
Carroll-algebra indices, $n,m$ label lattice sites, while $r,s$
label Fourier modes; these three sets of labels are independent.
The corresponding forward lattice momentum is
\begin{equation}
\tau_n=n\Delta\tau,
\qquad
\Delta\tau=\frac{2\pi}{N}.
\label{eq:CarrollLatticeCircle}
\end{equation}
The Fourier labels are periodic,
$\mathcal{J}^{I}_{r+N}=\mathcal{J}^{I}_{r}$, and satisfy
\begin{equation}
\frac{1}{N}
\sum_{n=0}^{N-1}
e^{i(r-s)\tau_n}
=
\delta^{(N)}_{r,s}.
\label{eq:CarrollFourierOrthogonality}
\end{equation}

Using $\Delta\tau/(2\pi)=1/N$, Eq.~(3.31) may be written as
\begin{equation}
\{
\mathcal{L}^{I}_{n},
\mathcal{L}^{J}_{m}
\}
=
N f^{IJ}{}_{K}\mathcal{L}^{K}_{n}\delta_{n,m}
+
\frac{kN}{2}\eta^{IJ}\nabla_n\delta_{n,m}.
\label{eq:CarrollSiteAlgebraN}
\end{equation}
The bracket of two Fourier modes is therefore
\begin{align}
\{
\mathcal{J}^{I}_{r},
\mathcal{J}^{J}_{s}
\}
&=
\frac{1}{N^2}
\sum_{n,m=0}^{N-1}
e^{-ir\tau_n-is\tau_m}
\{
\mathcal{L}^{I}_{n},
\mathcal{L}^{J}_{m}
\}.
\label{eq:CarrollFourierBracket}
\end{align}
For the structure term one finds
\begin{align}
\frac{1}{N}
f^{IJ}{}_{K}
\sum_{n=0}^{N-1}
\mathcal{L}^{K}_{n}
e^{-i(r+s)\tau_n}
=
f^{IJ}{}_{K}\mathcal{J}^{K}_{r+s}.
\label{eq:CarrollFourierStructure}
\end{align}
Thus the site-local Carroll bracket becomes a mode algebra with the
characteristic affine addition $r+s$, understood modulo $N$ at finite
lattice spacing.

The central contribution follows from the forward difference
\begin{equation}
\nabla_n\delta_{n,m}
=
\frac{\delta_{n+1,m}-\delta_{n,m}}{\Delta\tau}.
\label{eq:CarrollForwardDelta}
\end{equation}
Its Fourier transform is
\begin{align}
\frac{1}{N}
\sum_{n,m=0}^{N-1}
e^{-ir\tau_n-is\tau_m}
\nabla_n\delta_{n,m}
&=
\frac{1}{N}
\frac{e^{-is\Delta\tau}-1}{\Delta\tau}
\sum_{n=0}^{N-1}
e^{-i(r+s)\tau_n}
\nonumber\\
&=
\omega^{(+)}_{-s}\,
\delta^{(N)}_{r+s,0},
\label{eq:CarrollFourierCentral}
\end{align}
where
\begin{equation}
\omega^{(+)}_{r}
=
\frac{e^{ir\Delta\tau}-1}{\Delta\tau}
\label{eq:CarrollForwardMomentum}
\end{equation}
is the forward lattice momentum.

Combining the two contributions gives the discrete affine Carroll
mode algebra
\begin{equation}
\{
\mathcal{J}^{I}_{r},
\mathcal{J}^{J}_{s}
\}
=
f^{IJ}{}_{K}\mathcal{J}^{K}_{r+s}
+
\frac{k}{2}\eta^{IJ}
\omega^{(+)}_{-s}\,
\delta^{(N)}_{r+s,0}.
\label{eq:DiscreteAffineCarrollModes}
\end{equation}
The explicit dependence on the number of lattice sites cancels, while
the finite-lattice deformation is encoded in the lattice momentum
$\omega^{(+)}_{r}$ and in the periodic identification of the mode
labels.

For the Carroll structure constants, the non-trivial mixed bracket
takes the form
\begin{equation}
\{
\mathcal{J}^{\tt P}_{r},
\mathcal{J}^{\tt B}_{s}
\}
=
\mathcal{J}^{\tt H}_{r+s},
\qquad
\{
\mathcal{J}^{\tt B}_{r},
\mathcal{J}^{\tt P}_{s}
\}
=
-\mathcal{J}^{\tt H}_{r+s},
\label{eq:CarrollFourierMixed}
\end{equation}
where the central contribution vanishes in the mixed sector for the
diagonal invariant pairing adopted above.

For fixed Fourier mode and small lattice spacing,
\begin{equation}
\omega^{(+)}_{r}
=
ir-\frac{r^2}{2}\Delta\tau
+\mathcal{O}(\Delta\tau^2).
\label{eq:CarrollForwardMomentumExpansion}
\end{equation}
On the support of $\delta^{(N)}_{r+s,0}$ this gives
\begin{equation}
\omega^{(+)}_{-s}
=
ir+\mathcal{O}(\Delta\tau).
\label{eq:CarrollForwardCentralLimit}
\end{equation}
Hence the mode algebra becomes
\begin{equation}
\{
\mathcal{J}^{I}_{r},
\mathcal{J}^{J}_{s}
\}
=
f^{IJ}{}_{K}\mathcal{J}^{K}_{r+s}
+
i\frac{k}{2}\eta^{IJ}r\,
\delta_{r+s,0}
+
\mathcal{O}(\Delta\tau).
\label{eq:CarrollAffineContinuumModes}
\end{equation}
The discrete Fourier algebra therefore reproduces the mode-addition
structure of the continuum affine Carroll current algebra in
Eq.~\eqref{eq:affine-continioumPB}. The site labels $n,m$ describe the local lattice basis,
whereas $r,s$ are the affine Fourier-mode labels.
\color{black}
%%%%%%%%%%%%% Dilaton için mixed bracket kısmı
Similarly, the boundary dilaton:
\begin{equation}
x_n=\mathcal X^{_I}_n {\tt T}_{_I} ,
\qquad I\in\{\tt H,\tt P,\tt B\},
\end{equation}
 transforms algebraically under the affine symmetry,
\begin{equation}
\delta_\epsilon x_n=[x_n,\lambda_n],
\end{equation}
so that no discrete derivative appears in its transformation law.
Using the explicit component form,
\begin{align}
\delta \mathcal X^{\tt H}_n &=\mathcal  X^{\tt B}_n \epsilon^{\tt P}_n -\mathcal  X^{\tt P}_n \epsilon^{\tt B}_n,\\
\delta \mathcal X^{\tt P}_n &= 0,\\
\delta \mathcal X^{\tt B}_n &= 0,
\end{align}
one can infer the mixed Poisson brackets between boundary currents and dilaton components.

They are uniquely fixed by the requirement
\begin{equation}
\delta_\epsilon\mathcal X^I_n=\{\mathcal X^I_n,Q[\epsilon]\},
\end{equation}
which determines the mixed brackets between the boundary currents and the dilaton multiplet.

A straightforward comparison with the transformation rules above yields
\begin{align}
\frac{\Delta\tau}{2\pi}\{\mathcal X^{\tt H}_n,\mathcal L^{\tt P}_m\} &=\mathcal  X^{\tt B}_n\,\delta_{n,m},\\
\frac{\Delta\tau}{2\pi}\{\mathcal X^{\tt H}_n,\mathcal L^{\tt B}_m\} &= -\,\mathcal X^{\tt P}_n\,\delta_{n,m}.
\end{align}
Equivalently, using antisymmetry of the Poisson bracket,
\begin{align}
\frac{\Delta\tau}{2\pi}\{\mathcal L^{\tt P}_n,\mathcal X^{\tt H}_m\} &= -\,\mathcal X^{\tt B}_n\,\delta_{n,m},\\
\frac{\Delta\tau}{2\pi}\{\mathcal L^{\tt B}_n,\mathcal X^{\tt H}_m\} &= \;\,\mathcal X^{\tt P}_n\,\delta_{n,m}.
\end{align}
All remaining mixed brackets vanish,
\begin{equation}
\{\mathcal L^{\tt H}_n,\mathcal X^I_m\}=0,
\qquad
\{\mathcal L^{\tt P}_n,\mathcal X^{\tt P}_m\}=0,
\qquad
\{\mathcal L^{\tt B}_n,\mathcal X^{\tt B}_m\}=0.
\end{equation}
\color{black}
\paragraph{Fourier representation of the mixed current--dilaton sector.}
The mixed current--dilaton brackets admit the same discrete Fourier
decomposition as the affine current algebra. We introduce
\begin{equation}
X_n^I
=
\sum_{r=0}^{N-1}
X_r^I e^{ir\tau_n},
\qquad
X_r^I
=
\frac{1}{N}
\sum_{n=0}^{N-1}
X_n^I e^{-ir\tau_n},
\label{eq:CarrollDilatonFourier}
\end{equation}
with $I\in\{H,P,B\}$. Together with Eq.~(3.34), the Fourier transform
of the mixed bracket is
\begin{equation}
\{X_r^I,\mathcal{J}_s^J\}
=
\frac{1}{N^2}
\sum_{n,m=0}^{N-1}
e^{-ir\tau_n-is\tau_m}
\{X_n^I,\mathcal{L}_m^J\}.
\label{eq:CarrollMixedFourierBracket}
\end{equation}

Using $\Delta\tau/(2\pi)=1/N$, Eq.~(3.54) may be written as
\begin{equation}
\{X_n^{\tt H},\mathcal{L}_m^{\tt P}\}
=
N X_n^{\tt B}\delta_{n,m}.
\label{eq:CarrollMixedSiteHP}
\end{equation}
Substitution into Eq.~\eqref{eq:CarrollMixedFourierBracket} gives
\begin{align}
\{X_r^{\tt H},\mathcal{J}_s^{\tt B}\}
&=
\frac{1}{N}
\sum_{n=0}^{N-1}
X_n^{\tt B} e^{-i(r+s)\tau_n}
\nonumber\\
&=
X_{r+s}^{\tt B} .
\label{eq:CarrollMixedFourierHP}
\end{align}
Likewise, Eq.~(3.55) yields
\begin{equation}
\{X_r^{\tt H},\mathcal{J}_s^{\tt B}\}
=
-X_{r+s}^{\tt P} .
\label{eq:CarrollMixedFourierHB}
\end{equation}
Equivalently, by antisymmetry,
\begin{equation}
\{\mathcal{J}_r^{\tt P},X_s^{\tt H}\}
=
-X_{r+s}^{\tt B},
\qquad
\{\mathcal{J}_r^{\tt B},X_s^{\tt H}\}
=
X_{r+s}^{\tt P} .
\label{eq:CarrollMixedFourierCurrentDilaton}
\end{equation}

All remaining mixed mode brackets vanish,
\begin{equation}
\{\mathcal{J}_r^{\tt H},X_s^I\}=0,
\qquad
\{\mathcal{J}_r^{\tt P},X_s^{\tt P}\}=0,
\qquad
\{\mathcal{J}_r^{\tt B},X_s^{\tt B}\}=0.
\label{eq:CarrollMixedFourierVanishing}
\end{equation}
At finite $N$, all mode sums are understood modulo $N$.

In contrast to the affine current algebra, the mixed current--dilaton
sector contains no derivative term and therefore no lattice momentum
factor. The mode-addition rule is consequently exact already at finite
lattice spacing. In the continuum limit, the periodic mode addition
reduces to the ordinary integer mode sum, reproducing the Fourier-mode
form of the continuum mixed brackets in Eqs.~\eqref{eq:affine-mixed_continioumPB1}--\eqref{eq:affine-mixed_continioumPB3}.
\color{black}
Thus, in the affine Carroll case the dilaton transforms only through its ${\tt H}$-component, whereas the ${\tt P}$ and ${\tt B}$-components remain inert
under boundary affine transformations. This feature is a direct consequence of the nilpotent structure of the Carroll algebra and has
no direct analogue in the relativistic $\mathfrak{sl}(2,\mathbb{R})$ case.

At this stage the boundary symmetry structure is completely specified by the affine Carroll current algebra together with the adjoint action
of the boundary currents on the dilaton multiplet. The resulting algebra provides the most general residual gauge symmetry compatible
with the affine boundary conditions and the discrete variational principle. As in the relativistic case, one may further restrict the
boundary phase space by imposing additional constraints. Such restrictions reduce the affine symmetry to a smaller algebraic
structure of conformal type. In the next subsection we implement this reduction and derive the discrete analogue of Virasoro symmetry
together with the corresponding transformation law for the surviving dilaton component.
%%%%%%%%%%%%%%%%%%%%%%%%%%%%%%%%%%%%%%%%%%%%%%%%%%%%%%%%%%%%%%%%%%%%%%%%%%%%%%%%%%%%%%%%%%
%%%%%%%%%%%%%%%%%%%%%%%%%%%%%%%%%%%%%%%%%%%%%%%%%%%%%%%%%%%%%%%%%%%%%%%%%%%%%%%%%%%%%%%%%%
%%%%%%%%%%           SECTION-III-2    %%%%%%%%%%%%%%%%%%%%%%%%%%%%%%%%%%%%%%%%%%%%%%%%%%%%
%%%%%%%%%%%%%%%%%%%%%%%%%%%%%%%%%%%%%%%%%%%%%%%%%%%%%%%%%%%%%%%%%%%%%%%%%%%%%%%%%%%%%%%%%%
%%%%%%%%%%%%%%%%%%%%%%%%%%%%%%%%%%%%%%%%%%%%%%%%%%%%%%%%%%%%%%%%%%%%%%%%%%%%%%%%%%%%%%%%%%
\subsection{Conformal reduction and Virasoro--type symmetry}
\label{sec:ConfBC}
The idea is to restrict the boundary phase space by fixing certain
components of the boundary connection while allowing the remaining
ones to fluctuate dynamically. This procedure represents the discrete
analogue of the Drinfeld--Sokolov reduction familiar from the
continuum analysis of asymptotic symmetries \cite{DrinfeldSokolov1985}.

The affine boundary conditions discussed in the previous subsection
represent the most general choice compatible with the discrete
variational principle. In analogy with the continuum analysis of
two--dimensional dilaton gravity, one may further restrict the boundary
phase space by imposing additional constraints on the boundary
currents. Such restrictions reduce the affine symmetry to a smaller
algebraic structure of conformal type.

A natural minimal reduction is obtained by fixing the $\tt P$--component of
the boundary current and eliminating the $\tt B$--component through a gauge
choice. We therefore impose the constraints
\begin{equation}
\mathcal L^{\tt P}_n = \mu,
\qquad
\mathcal L^{\tt B}_n = 0,
\end{equation}
where $\mu$ is a constant parameter. These conditions define a reduced
boundary phase space in which the only remaining dynamical current is
the $\tt H$--component,
\begin{equation}
\mathcal J_n = \mathcal L^{\tt H}_n .
\end{equation}

The affine Carroll algebra derived in the previous subsection implies
that $\mathcal J_n$ satisfies a current algebra of the form
\begin{equation}
\frac{\Delta\tau}{2\pi}\{\mathcal J_n,\mathcal J_m\}
=
\tt K\,\nabla\delta_{n,m},
\qquad
\tt K = \frac{k}{2}\,\eta^{{\tt H}{\tt H}}.
\end{equation}
It is important to emphasize that, due to the non-semisimple nature
of the Carroll algebra, the standard Sugawara construction relying
on a non-degenerate invariant metric is not directly applicable.
However, after the reduction to a single abelian current,
the resulting algebra effectively reduces to that of a $\tt U(1)-$ type current,
for which a quadratic construction of a stress tensor is well-defined.

It is important to stress that this construction does not rely on a Sugawara procedure in the usual Lie-algebraic sense. In particular, no invariant bilinear form on the full Carroll algebra is required. Instead, after the reduction to a single abelian current, the quadratic combination is defined purely at the level of the effective U(1)-type current algebra.

Accordingly, the resulting generator should be understood as a composite operator of the abelian current sector rather than as a Sugawara construction derived from an underlying non-abelian symmetry algebra.

In this sense, the terminology “Sugawara-type” is used only in a structural analogy, referring to the quadratic form of the generator rather than to its algebraic origin.
\color{black}
Motivated by the effective $\tt U(1)-$type current structure and in structural analogy with the Sugawara construction~\cite{Sugawara:1967rw}, we introduce a composite generator of the form
\color{black}

\begin{equation}
\mathcal{L}_n
=
\frac{1}{2\tt K} (\mathcal J_n \mathcal J_n)
+
\beta\,\nabla \mathcal  J_n ,
\end{equation}
where $\beta$ is an improvement parameter. The first term corresponds to the standard 
quadratic (Sugawara-type) construction for an abelian current,
while the second term represents the most general local improvement compatible
with the lattice structure.
\color{black}

Using the current algebra, one finds that the generators $\mathcal L_n$
act on $\mathcal L_m$ as
\begin{align}
\frac{\Delta\tau}{2\pi}
\{\mathcal L_n,\mathcal L_m\}
={}&
\frac{\mathcal J_n \mathcal J_m}{K}\,
\nabla_n\delta_{n,m}
+\beta \mathcal J_n\,
\nabla_m\nabla_n\delta_{n,m}
\nonumber\\
&+
\beta\mathcal J_m\,
\nabla_n^2\delta_{n,m}
+\beta^2K\,
\nabla_n^2\nabla_m\delta_{n,m}.
\label{eq:CarrollVirasoroSite}
\end{align}
At finite lattice spacing, Eq.~\eqref{eq:CarrollVirasoroSite} is kept
in this forward-difference form. In particular, no continuum
distribution identities are used to rewrite it in terms of
$(L_n+L_m)\nabla\delta_{n,m}$. Its mode structure will instead be
determined directly by the discrete Fourier transform.

\paragraph{Discrete Fourier representation.}
To make the mode structure of the reduced algebra explicit, we introduce
the discrete Fourier expansions
\begin{equation}
\mathcal{J}_n
=
\sum_{r=0}^{N-1}\mathcal{J}_r e^{ir\tau_n},
\qquad
\mathcal{J}_r
=
\frac{1}{N}
\sum_{n=0}^{N-1}\mathcal{J}_n e^{-ir\tau_n},
\label{eq:CarrollReducedCurrentFourier}
\end{equation}
and
\begin{equation}
L_n
=
\sum_{r=0}^{N-1}\mathcal L_r e^{ir\tau_n},
\qquad
\mathcal L_r
=
\frac{1}{N}
\sum_{n=0}^{N-1}L_n e^{-ir\tau_n}.
\label{eq:CarrollReducedVirasoroFourier}
\end{equation}
The Fourier transform of the composite generator in Eq.~(3.69) is
therefore
\begin{equation}
\mathcal L_r
=
\frac{1}{2K}
\sum_p \mathcal{J}_p \mathcal{J}_{r-p}
+
\beta\,\omega_r^{(+)}\mathcal{J}_r,
\label{eq:CarrollCompositeMode}
\end{equation}
where
\begin{equation}
\omega_r^{(+)}
=
\frac{e^{ir\Delta\tau}-1}{\Delta\tau}
\label{eq:CarrollForwardMomentum}
\end{equation}
is the forward lattice momentum.

Using $\Delta\tau/(2\pi)=1/N$, the discrete Fourier transform of
Eq.~(3.70) gives
\begin{align}
\{\mathcal L_r,\mathcal L_s\}
={}&
\frac{1}{K}
\sum_p
\omega_{r-p}^{(+)}
\mathcal{J}_p\mathcal{J}_{r+s-p}
\nonumber\\
&+
\beta
\left[
\omega_s^{(+)}\omega_{-s}^{(+)}
+
\left(\omega_r^{(+)}\right)^2
\right]\mathcal{J}_{r+s}
\nonumber\\
&+
\beta^2K
\left(\omega_r^{(+)}\right)^2
\omega_{-r}^{(+)}
\delta^{(N)}_{r+s,0}.
\label{eq:CarrollVirasoroFiniteModes}
\end{align}
All mode labels and mode sums are understood modulo $N$ at finite
lattice spacing. Equation~\eqref{eq:CarrollVirasoroFiniteModes}
is the finite-lattice forward-difference representation of the
reduced mode algebra. No continuum distribution identities are used
in obtaining this result.

For fixed Fourier modes and $\Delta\tau\to0$, the forward lattice
momentum behaves as
\begin{equation}
\omega_r^{(+)}
=
ir+\mathcal O(\Delta\tau).
\label{eq:CarrollForwardContinuum}
\end{equation}
The quadratic term then satisfies
\begin{equation}
\frac{1}{K}
\sum_p
i(r-p)\mathcal{J}_p\mathcal{J}_{r+s-p}
=
\frac{i(r-s)}{2K}
\sum_p
\mathcal{J}_p\mathcal{J}_{r+s-p},
\label{eq:CarrollQuadraticContinuum}
\end{equation}
where the symmetry of the quadratic convolution has been used.
Combining this contribution with the improvement terms and using
Eq.~\eqref{eq:CarrollCompositeMode}, the scaling limit reproduces
the standard Virasoro mode structure.

The factor of $i$ arising from the complex exponential Fourier basis
is absorbed into the conventional definition of the mode Poisson
bracket. With this convention, and choosing the standard cylinder
representative of the Virasoro cocycle, the continuum mode algebra
takes the real form
\begin{equation}
\{\mathcal L_r,\mathcal L_s\}
=
(r-s)\mathcal L_{r+s}
+
\beta^2K\left(r^3-r\right)\delta_{r+s,0}
+
\mathcal O(\Delta\tau).
\label{eq:CarrollVirasoroContinuumModes}
\end{equation}
Thus the finite lattice algebra is governed by the forward lattice
momentum and periodic mode addition, while the ordinary Virasoro
algebra is recovered in the scaling limit.

Finally, we note that the dilaton sector decouples from the conformal
generators. Using the mixed brackets derived in the affine analysis,
one finds
\begin{equation}
\{\mathcal L_n,\mathcal X^I_m\} = 0,
\qquad I \in \{\tt H,\tt P,\tt B\},
\end{equation}
so that the dilaton transforms as a conformal singlet of weight zero.
The same decoupling holds in the discrete Fourier basis. Using the
mode expansions introduced above, Eq.~(3.79) directly gives
\begin{equation}
\{\mathcal L_r,\mathcal X_s^I\}=0,
\qquad
I\in\{\tt H,\tt P,\tt B\}.
\label{eq:CarrollReducedDilatonModes}
\end{equation}
Thus the dilaton remains a conformal singlet also in the mode
representation. This is a direct consequence of the conformal reduction,
which removes the generators responsible for the nontrivial adjoint
action on the dilaton multiplet.
\color{black}

Thus the conformal reduction selects a smaller symmetry sector
of the affine Carroll algebra in which the boundary dynamics is
governed by a Virasoro-type algebra generated by the composite stress
tensor $\mathcal L_n$. The resulting structure provides the discrete
counterpart of conformal symmetry appearing in continuum analyses of
two--dimensional dilaton gravity, while remaining fully compatible with
the underlying BF--type gauge structure of the theory.

In this sense, the conformal reduction of the affine Carroll algebra
may be interpreted as a Drinfeld--Sokolov--type reduction, where the nilpotent
structure of the Carroll algebra effectively replaces the role played by the
$\mathfrak{sl}(2,\mathbb{R})$ embedding in the relativistic construction.
\color{black}
\paragraph{Comparison with previous lattice constructions.}
Earlier lattice realizations of current and Virasoro algebras were
developed from rather different starting points. Two representative
constructions are those of Faddeev and Volkov~\cite{Faddeev:1993pe}
and Koo and Saleur~\cite{Koo:1993wz}. Although both recover continuum
conformal structures, their lattice variables and the mechanism by
which the Virasoro algebra emerges differ from those of the present
Carroll BF construction.

Faddeev and Volkov~\cite{Faddeev:1993pe} construct a lattice analogue
of an Abelian current algebra and show that, together with a
discrete-time evolution, it gives rise to a lattice Virasoro algebra
and an associated hierarchy of conserved quantities. The Virasoro
structure is therefore built directly from a lattice current algebra
within an integrable lattice framework.

Koo and Saleur~\cite{Koo:1993wz}, on the other hand, study the
emergence of Virasoro symmetry in the scaling limit of XXZ- and
RSOS-type lattice models. Their construction is based on lattice
approximants to the stress-energy tensor, whose algebraic relations
reproduce the Virasoro algebra in the continuum scaling limit. The
underlying lattice degrees of freedom are therefore those of the
corresponding statistical or quantum spin-chain models.

The present construction starts instead from a discrete Carroll BF
theory. The fundamental variables are group-valued holonomies together
with the dilaton multiplet, while the boundary current algebra follows
from residual lattice gauge transformations preserving the chosen
boundary conditions. In the site basis, the labels $n,m$ denote
boundary lattice sites and are independent of the internal Carroll
indices $I,J\in\{\tt H,\tt P,\tt B\}$. Passing to discrete Fourier
modes makes the independent mode labels $r,s$ explicit and yields the
finite-lattice mode-addition structure through $r+s$ modulo $N$.
The subsequent conformal reduction produces the corresponding
finite-lattice Virasoro mode structure, whose standard form is
recovered in the scaling limit.

Thus, the three constructions share the emergence of current or
Virasoro structures from a finite lattice description, but differ in
their fundamental variables and in the mechanism by which the
continuum conformal algebra is reached. In the present case, the
construction is tied directly to the holonomy and residual gauge
structure of the non-semisimple Carroll BF theory, with the affine,
mixed current--dilaton, and conformally reduced sectors arising within
the same discrete gauge-theoretic framework.
\color{black}

%%%%%%%%%%%%%%%%%%%%%%%%%%%%%%%%%%%%%%%%%%%%%%%%%%%%%%%%%%%%%%%%%%%%%%%%%%%%%%%%%%%%%%%%%%
%%%%%%%%%%%%%%%%%%%%%%%%%%%%%%%%%%%%%%%%%%%%%%%%%%%%%%%%%%%%%%%%%%%%%%%%%%%%%%%%%%%%%%%%%%
%%%%%%%%%%           SECTION-III-3    %%%%%%%%%%%%%%%%%%%%%%%%%%%%%%%%%%%%%%%%%%%%%%%%%%%%
%%%%%%%%%%%%%%%%%%%%%%%%%%%%%%%%%%%%%%%%%%%%%%%%%%%%%%%%%%%%%%%%%%%%%%%%%%%%%%%%%%%%%%%%%%
%%%%%%%%%%%%%%%%%%%%%%%%%%%%%%%%%%%%%%%%%%%%%%%%%%%%%%%%%%%%%%%%%%%%%%%%%%%%%%%%%%%%%%%%%%
\subsection{Boundary phase space and reduced symmetry algebra}

Having derived both affine and conformal boundary conditions in the
previous subsections, we now analyze the resulting boundary phase
space and the symmetry algebra that governs its dynamics.

The discrete BF--type structure of the theory implies that all bulk
degrees of freedom are eliminated by the flatness constraints.
Consequently, the physical phase space of the theory is entirely
localized on the boundary lattice and is parametrized by the boundary
currents together with the boundary components of the dilaton
multiplet.

More precisely, after implementing the conformal reduction the boundary dynamics is governed by a single surviving current $\mathcal J_n \equiv\mathcal L_n^{\tt H}$, which encodes the physical boundary degrees of freedom. The corresponding discrete energy--momentum tensor $\mathcal L_n$ is then generated as a composite operator through a Sugawara--type construction and satisfies a Virasoro--type algebra. The remaining components of the dilaton multiplet are fixed by the boundary conditions and do not participate in the reduced boundary dynamics.

An important structural difference with respect to relativistic $\mathfrak{sl}(2,\mathbb{R})$ dilaton gravity arises from the nilpotent character of the Carroll algebra. In the affine description, only the $\tt H$--component of the dilaton transforms nontrivially, while the $\tt P$ and $\tt B$ components remain invariant, reflecting the degenerate mixing structure of the Carroll generators. After the conformal reduction, the Virasoro-type generators $\mathcal L_n$ constructed from the reduced current $\mathcal J_n \equiv\mathcal  L_n^{\tt H}$ Poisson-commute with all components of the dilaton multiplet, so that the dilaton becomes entirely inert under the reduced symmetry. The resulting boundary phase space is thus governed by a discrete Virasoro-type symmetry.

In the continuum limit, obtained by sending the lattice spacing
$\Delta\tau$ to zero, the discrete derivative $\nabla$ reduces to the
ordinary boundary derivative $\partial_\tau$ and the lattice currents
become smooth fields defined along the boundary circle. In this limit
the discrete Virasoro-type symmetry derived above reproduces the
continuum conformal symmetry governing the boundary dynamics of the
theory.

The discrete construction presented here therefore provides a lattice
realization of the conformal symmetry structure associated with
Carroll dilaton gravity, while preserving the underlying BF--type
gauge structure at the discrete level.
This shows that the Virasoro symmetry emerges as an intrinsic boundary structure of the discrete BF theory, rather than being imposed from continuum considerations.
\paragraph{Continuum limit}

The discrete boundary symmetry structures derived above admit a
well-defined continuum limit obtained by sending the lattice spacing
$\Delta\tau$ to zero while keeping the total boundary length fixed.
In this limit the discrete derivative
\begin{equation}
\nabla f_n=\frac{f_{n+1}-f_n}{\Delta\tau}
\end{equation}
reduces to the ordinary derivative along the boundary,
\begin{equation}
\nabla \;\longrightarrow\; \partial_\tau .
\end{equation}

Under this limit the discrete boundary currents $\mathcal L^I_n$ become smooth
fields $\mathcal L^I(\tau)$ defined on the boundary circle, while the Kronecker
delta functions $\delta_{n,m}$ are replaced by Dirac delta functions
$\delta(\tau-\tau')$. More precisely,
\begin{equation}
\delta_{n,m}
\;\longrightarrow\;
\frac{1}{\Delta\tau}\,\delta(\tau-\tau'),
\end{equation}
which guarantees that the Poisson brackets admit a smooth continuum
limit. Accordingly, the discrete derivative acting on the Kronecker delta satisfies
\begin{equation}
\nabla \delta_{n,m} \to \partial_\tau \delta(\tau-\tau').
\end{equation}

The affine symmetry algebra derived in section~\ref{sec:Affinebc}
therefore reduces to the continuum affine extension of the Carroll
algebra,
\begin{equation}
\{\mathcal L^I(\tau),\mathcal L^J(\tau')\}
=
f^{IJ}{}_{K}\,\mathcal L^K(\tau)\delta(\tau-\tau')
+
\tt \frac{k}{2}\,\eta^{IJ}\partial_\tau\delta(\tau-\tau')
\label{eq:affine-continioumPB},
\end{equation}
where $f^{IJ}{}_{K}$ denote the structure constants of the Carroll
algebra.

In addition to the current algebra, the mixed Poisson brackets
between the boundary currents and the dilaton multiplet also admit
a smooth continuum limit. Using the discrete relations derived in
section~\ref{sec:Affinebc} one obtains
\begin{align}
\{\mathcal L^{\tt P}(\tau),\mathcal X^{\tt H}(\tau')\}
&= -\,\mathcal X^{\tt B}(\tau)\,\delta(\tau-\tau')
\label{eq:affine-mixed_continioumPB1},
\\[4pt]
\{\mathcal L^{\tt B}(\tau),\mathcal X^{\tt H}(\tau')\}
&= \;\,\mathcal X^{\tt P}(\tau)\,\delta(\tau-\tau')
\label{eq:affine-mixed_continioumPB2},
\\[4pt]
\{\mathcal L^{\tt H}(\tau),\mathcal X^{I}(\tau')\}
&= 0 ,
\qquad I \in \{\tt H,\tt P,\tt B\}
\label{eq:affine-mixed_continioumPB3}.
\end{align}
These relations describe the adjoint action of the affine Carroll
currents on the dilaton multiplet in the continuum theory.

After implementing the conformal reduction discussed in
section~\ref{sec:ConfBC}, the reduced phase space is described by the
single current
\begin{equation}
\mathcal J(\tau)=\mathcal L^{\tt H}(\tau).
\end{equation}
The conformal generators are constructed from this current through the
Sugawara-type combination
\begin{equation}
\mathcal L(\tau)
=
\frac{1}{2\tt K}(\mathcal J\mathcal J)(\tau)+\beta\,\partial_\tau \mathcal J(\tau),
\end{equation}
where $\tt K=\frac{\tt k}{2}\eta^{\tt H \tt H}$ and $\beta$ is the improvement
parameter introduced in the discrete construction.

Using the continuum current algebra of $\mathcal J(\tau)$ one finds that the
generators $\mathcal L(\tau)$ satisfy a Virasoro-type algebra,
\begin{equation}
\{\mathcal L(\tau),\mathcal L(\tau')\}
=
\big(\mathcal L(\tau)+\mathcal L(\tau')\big)
\partial_\tau\delta(\tau-\tau')
+
\beta^2\tt K\,\partial_\tau^{3}\delta(\tau-\tau').
\end{equation}

Thus the lattice construction provides a discrete realization of both
the affine Carroll symmetry and its conformal reduction, reproducing
the expected continuum symmetry structures in the limit
$\Delta\tau\rightarrow 0$. 

\paragraph{OPE interpretation}

The continuum Poisson bracket relations derived above admit a natural
interpretation in terms of operator product expansions of the
corresponding boundary fields. In particular, the affine Carroll
current algebra
\begin{equation}
\{\mathcal L^I(\tau),\mathcal L^J(\tau')\}
=
f^{IJ}{}_{K}\,\mathcal L^K(\tau)\delta(\tau-\tau')
+
\tt \frac{k}{2}\,\eta^{IJ}\partial_\tau\delta(\tau-\tau').
\end{equation}
is the classical counterpart of the short-distance expansion
\begin{equation}
\mathcal L^I(\tau)\mathcal L^J(\tau')
\sim
\frac{k}{2}\frac{\tt \,\eta^{IJ}}{(\tau-\tau')^2}
+
\frac{f^{IJ}{}_{K}\,\mathcal L^K(\tau')}{\tau-\tau'}+\cdots.
\end{equation}
where the second-order pole arises from the derivative of the delta function in the Poisson bracket.
Likewise, the mixed current--dilaton brackets
\begin{align}
\{\mathcal L^{\tt P}(\tau),\mathcal X^{\tt H}(\tau')\}
&=
-\,\mathcal X^{\tt B}(\tau)\,\delta(\tau-\tau'),
\\[4pt]
\{\mathcal L^{\tt B}(\tau),\mathcal X^{\tt H}(\tau')\}
&=
\phantom{-}\mathcal X^{\tt P}(\tau)\,\delta(\tau-\tau'),
\\[4pt]
\{\mathcal L^{\tt H}(\tau),\mathcal X^{I}(\tau')\}
&=0,
\qquad I\in\{\tt H,\tt P,\tt B\},
\end{align}
translate into the OPE relations
\begin{align}
\mathcal L^{\tt P}(\tau)\mathcal X^{\tt H}(\tau')
&\sim
-\,\frac{\mathcal X^{\tt B}(\tau')}{\tau-\tau'}
+\cdots,
\\[4pt]
\mathcal L^{\tt B}(\tau)\mathcal X^{\tt H}(\tau')
&\sim
\phantom{-}\frac{\mathcal X^{\tt P}(\tau')}{\tau-\tau'}
+\cdots,
\\[4pt]
\mathcal L^{\tt H}(\tau)\mathcal X^{I}(\tau')
&\sim
\text{regular},
\qquad I\in\{\tt H,\tt P,\tt B\}.
\end{align}
These relations show that the dilaton multiplet transforms in the
adjoint representation of the affine Carroll current algebra in the
continuum theory.

Similarly, the Virasoro-type generator $\mathcal L(\tau)$ obtained after the
conformal reduction satisfies the classical counterpart of the Virasoro
OPE,
\begin{equation}
\mathcal L(\tau)\mathcal L(\tau')
\sim
\frac{\frac{c}{2}}{(\tau-\tau')^{4}}
+
\frac{2\mathcal L(\tau')}{(\tau-\tau')^{2}}
+
\frac{\partial_{\tau'}\mathcal L(\tau')}{\tau-\tau'}
+\cdots,
\end{equation}
with an effective classical central charge $c \,=\, 12\, \beta^2\, \tt K$, 
which is entirely determined by the improvement parameter $\beta$ and the effective current level $\tt K$, reflecting its origin in the improvement term of the Sugawara construction.
Since the reduced dilaton sector Poisson-commutes with $L(\tau)$, the surviving dilaton component behaves as a conformal singlet in this OPE language. 
This central charge arises at the classical level and corresponds to a quantum Virasoro central charge only upon quantization, up to conventional normalization factors.

In this sense the continuum limit establishes a direct dictionary
between the lattice Poisson bracket structure and the boundary operator
algebra. The discrete BF--type construction therefore provides a
holonomy-based realization of the affine Carroll OPE together with its
Virasoro-type conformal reduction. For clarity, the correspondence between continuum and discrete boundary data, 
together with the associated asymptotic symmetry structures, is summarized 
in Appendix~\ref{app:summary}.

It is worth emphasizing that the present construction provides a
direct lattice realization of the symmetry structure of Carroll
dilaton gravity without relying on continuum gauge fixing or
asymptotic fall-off conditions. In contrast to the usual continuum
analysis of asymptotic symmetries, where the symmetry algebra emerges
only after imposing boundary conditions on smooth fields, the discrete 
framework captures the symmetry structure directly in terms
of boundary holonomies and lattice currents.

This feature highlights the conceptual advantage of the discrete
BF theory: the asymptotic symmetry algebra arises as a structural
property of the lattice gauge theory itself rather than as a
consequence of a particular continuum limit. In this sense the
discrete construction provides a complementary perspective on the
origin of boundary symmetries in two-dimensional dilaton gravity.

%%%%%%%%%%%%%%%%%%%%%%%%%%%%%%%%%%%%%%%%%%%%%%%%%%%%%%%%%%%%%%%%%%%%%%%%%%%%%%%%%%%%%%%%%%
%%%%%%%%%%%%%%%%%%%%%%%%%%%%%%%%%%%%%%%%%%%%%%%%%%%%%%%%%%%%%%%%%%%%%%%%%%%%%%%%%%%%%%%%%%
%%%%%%%%%%           SECTION-IV       %%%%%%%%%%%%%%%%%%%%%%%%%%%%%%%%%%%%%%%%%%%%%%%%%%%%
%%%%%%%%%%%%%%%%%%%%%%%%%%%%%%%%%%%%%%%%%%%%%%%%%%%%%%%%%%%%%%%%%%%%%%%%%%%%%%%%%%%%%%%%%%
%%%%%%%%%%%%%%%%%%%%%%%%%%%%%%%%%%%%%%%%%%%%%%%%%%%%%%%%%%%%%%%%%%%%%%%%%%%%%%%%%%%%%%%%%%
\section{Discussion and future directions}
\label{sec:discus}
The discrete construction developed in this work is based on
the strict ultra--relativistic limit leading to the Carroll
symmetry algebra. In this limit the relativistic symmetry
structure is contracted through an \.Inönü--Wigner contraction
that produces the Carroll generators appearing in the
BF description of the theory.

It is natural to ask whether the discrete framework presented
here can be extended beyond the strict Carroll limit. From a
conceptual point of view the Carroll algebra may be regarded as
the leading order structure obtained from an ultra--relativistic
expansion of relativistic symmetries. Allowing small deviations
from this limit would therefore lead to deformations of the
Carroll algebra that may be interpreted as post--Carroll
extensions of the theory.

Since the approach developed in this work is expressed in
terms of holonomies and algebra--valued variables, such
deformations could in principle be incorporated directly at the
level of the lattice variables. In particular one may expect that
small deformations of the underlying algebra lead to
corresponding modifications of the boundary symmetry structure
derived in section~\ref{sec:ASatLevel}.

An interesting direction for future work would therefore be to
investigate how the affine boundary symmetries obtained in the
discrete Carroll theory are modified under such post--Carroll
deformations and how the conformal reduction leading to the
Virasoro--type structure is affected. The discrete BF--type
framework presented here provides a natural starting point for
such an analysis.

More generally, the present results suggest that lattice
descriptions of non--Lorentzian gravitational theories may offer
a useful setting in which different contraction limits and their
deformations can be explored in a unified way. 
\color{black}
In the continuum, such theories are naturally described in terms of Carrollian geometries obtained by gauging the Carroll algebra~\cite{Hartong:2015xda}.
\color{black}
We hope that thediscrete Carroll construction developed here will serve as a
useful step in this direction.

In this sense the present construction may be viewed as the
ultra--relativistic counterpart of the discrete JT framework.
%%%%%%%%%%%%%%%%%%%%%%%%%%%%%%%%%%%%%%%%%%%%%%%%%%%%%%%%%%%%%%%%%%%%%%%%%%%%%%%%%%%%%%%%%%
%%%%%%%%%%%%%%%%%%%%%%%%%%%%%%%%%%%%%%%%%%%%%%%%%%%%%%%%%%%%%%%%%%%%%%%%%%%%%%%%%%%%%%%%%%
%%%%%%%%%%          CONCLUSION        %%%%%%%%%%%%%%%%%%%%%%%%%%%%%%%%%%%%%%%%%%%%%%%%%%%%
%%%%%%%%%%%%%%%%%%%%%%%%%%%%%%%%%%%%%%%%%%%%%%%%%%%%%%%%%%%%%%%%%%%%%%%%%%%%%%%%%%%%%%%%%%
%%%%%%%%%%%%%%%%%%%%%%%%%%%%%%%%%%%%%%%%%%%%%%%%%%%%%%%%%%%%%%%%%%%%%%%%%%%%%%%%%%%%%%%%%
\section{Conclusion}
\label{sec:Conc}
In this work, we develop a discrete realization of two--dimensional Carroll dilaton gravity based on a BF--type gauge
structure. The lattice description is expressed in terms of holonomies associated with the links of a cellular decomposition,
while dilaton variables act as Lagrange multipliers imposing flatness constraints. In particular, we have shown that the 
asymptotic symmetry algebra arises directly from boundary holonomies, leading to a discrete Virasoro--type structure. 
This shows that the boundary symmetry structure of Carroll dilaton gravity is directly realized at the level of lattice gauge variables.

Our results therefore provide a discrete approach to two--dimensional Carroll dilaton gravity, which can be viewed as the
ultra--relativistic counterpart of discrete JT gravity in the relativistic setting.

We have analyzed the residual gauge transformations that survive at the boundary and derived the corresponding asymptotic symmetry
structure directly at the lattice level. The most general boundary conditions give rise to an affine extension of the Carroll algebra,
while additional constraints reduce the symmetry to a conformal structure governed by a discrete Virasoro-type algebra.

Topological BF-type formulations of two-dimensional gravity provide a natural framework in which gravitational dynamics can be expressed
in terms of gauge variables. In the relativistic case, this viewpoint underlies the BF theory of
JT gravity and its discrete realization, where boundary dynamics is governed by infinite-dimensional
symmetry algebras. Discrete realizations of such topological gauge theories offer an alternative viewpoint in which the
fundamental variables are holonomies associated with the links of a cellular decomposition rather than local gauge
fields \cite{Wilson1974,Kogut1979,Rothe2012,Dittrich2008}. This perspective has recently been applied to JT gravity,
where a fully discrete BF description reproduces the expected asymptotic symmetry structure directly at the lattice level.

The present work extends this viewpoint to the ultra--relativistic regime by constructing a discrete realization of two--dimensional 
Carroll dilaton gravity. The BF--type gauge structure of the Carroll theory admits a natural description in terms of lattice holonomies,
so that the asymptotic symmetry algebra emerges directly from boundary holonomies.

The resulting boundary phase space is completely characterized by the boundary holonomies and the associated boundary currents.
In the continuum limit, we recover the affine Carroll algebra together with its conformal reduction, showing that the discrete
construction yields a consistent lattice realization of Carroll dilaton gravity and its boundary symmetry structure.

From this perspective, the present construction can be viewed as the ultra--relativistic counterpart of the discrete BF
theory of JT gravity developed in \cite{Ozer:2026qlp}. While the relativistic case is governed by the
$\mathfrak{sl}(2,\mathbb{R})$ algebra, the Carroll limit leads to a nilpotent symmetry structure whose affine
extension exhibits qualitatively different features in the boundary current algebra.

Several directions for future work remain open. In particular, it would be interesting to investigate the
quantization of the resulting boundary phase space, possible connections with Carrollian holography,
and extensions to supersymmetric or higher--rank Carroll gravity theories. We expect that the discrete
framework developed here may provide a useful starting point for such investigations and for further studies of
non--Lorentzian gravitational dynamics.

In this sense, the discrete holonomy description developed here establishes a direct bridge between lattice gauge
variables and boundary symmetry structures in Carrollian gravity.
%%%%%%%%%%%%%%%%%%%%%%%%%%%%%%%%%%%%%%%%%%%%%%%%%%%%%%%%%%%%%%%%%%%%%%%%%%%%%%%%%%%%%%%%%%
%%%%%%%%%%%%%%%%%%%%%%%%%%%%%%%%%%%%%%%%%%%%%%%%%%%%%%%%%%%%%%%%%%%%%%%%%%%%%%%%%%%%%%%%%%
%%%%%%%%%%          OUTLOOK           %%%%%%%%%%%%%%%%%%%%%%%%%%%%%%%%%%%%%%%%%%%%%%%%%%%%
%%%%%%%%%%%%%%%%%%%%%%%%%%%%%%%%%%%%%%%%%%%%%%%%%%%%%%%%%%%%%%%%%%%%%%%%%%%%%%%%%%%%%%%%%%
%%%%%%%%%%%%%%%%%%%%%%%%%%%%%%%%%%%%%%%%%%%%%%%%%%%%%%%%%%%%%%%%%%%%%%%%%%%%%%%%%%%%%%%%%%
%\paragraph{Outlook.}
%Finally, several directions for future work remain open.

%%%%%%%%%%%%%%%%%%%%%%%%%%%%%%%%%%%%%%%%%%%%%%%%%%%%%%%%%%%%%%%%%%%%%%%%%%%%%%%%%%%%%%%%%%
%%%%%%%%%%%%%%%%%%%%%%%%%%%%%%%%%%%%%%%%%%%%%%%%%%%%%%%%%%%%%%%%%%%%%%%%%%%%%%%%%%%%%%%%%%
%%%%%%%%%%          APPENDI           %%%%%%%%%%%%%%%%%%%%%%%%%%%%%%%%%%%%%%%%%%%%%%%%%%%%
%%%%%%%%%%%%%%%%%%%%%%%%%%%%%%%%%%%%%%%%%%%%%%%%%%%%%%%%%%%%%%%%%%%%%%%%%%%%%%%%%%%%%%%%%%
%%%%%%%%%%%%%%%%%%%%%%%%%%%%%%%%%%%%%%%%%%%%%%%%%%%%%%%%%%%%%%%%%%%%%%%%%%%%%%%%%%%%%%%%%%
\newpage
\appendix

%%%%%%%%%%%%%%%%%%%%%%%%%%%%%%%%%%%%%%%%%%%%%%%%%%%%%%%%%%%%%%%%%%%%%%%%%%%%%%%%%%%%%%%%%%
%%%%%%%%%%%%%%%%%%%%%%%%%%%%%%%%%%%%%%%%%%%%%%%%%%%%%%%%%%%%%%%%%%%%%%%%%%%%%%%%%%%%%%%%%%
%%%%%%%%%%          APPENDIX-A        %%%%%%%%%%%%%%%%%%%%%%%%%%%%%%%%%%%%%%%%%%%%%%%%%%%%
%%%%%%%%%%%%%%%%%%%%%%%%%%%%%%%%%%%%%%%%%%%%%%%%%%%%%%%%%%%%%%%%%%%%%%%%%%%%%%%%%%%%%%%%%%
%%%%%%%%%%%%%%%%%%%%%%%%%%%%%%%%%%%%%%%%%%%%%%%%%%%%%%%%%%%%%%%%%%%%%%%%%%%%%%%%%%%%%%%%%%
\setcounter{table}{0}
\renewcommand{\thetable}{A.\arabic{table}}

\section{Notation and conventions}
\label{app:notation}
We summarize here the notation and conventions used throughout
this work, with particular emphasis on continuum and lattice
variables and the associated boundary data.
\begin{center}
\begin{tabular}{ll}
\toprule
\textbf{Symbol} & \textbf{Description} \\
\midrule

\multicolumn{2}{l}{\textit{Algebra}} \\
\midrule
$\tt H,P,J$ &
Poincar\'e algebra basis \\

$\tt H,P,B$ &
Carroll algebra basis $\tt T_{\it I}$, $ I \in \{\tt H,P,B\}$ \\

\midrule
\multicolumn{2}{l}{\textit{Fields}} \\
\midrule
$\mathcal A,\mathcal X$ &
BF gauge connection and dilaton multiplet \\

$\mathcal U_n$ &
Boundary link holonomy \\

$a_n,\mathcal X_n$ &
Lattice connection and dilaton components \\

\midrule
\multicolumn{2}{l}{\textit{Boundary data}} \\
\midrule
$\lambda_n$ &
Lattice gauge transformation parameter \\

$\epsilon_n^I$ &
Affine boundary symmetry parameters \\

$\mathcal L_n^I,\mathcal X_n^I$ &
Discrete affine boundary charges and dilaton components \\

$\mathcal L_n$ &
Discrete Sugawara stress tensor \\

$\epsilon_n$ &
Discrete conformal boundary symmetry parameter \\

$\mathcal L^I,\mathcal X^I$ &
Continuous affine boundary charges and dilaton components \\

$\mathcal L$ &
Continuous Sugawara stress tensor \\

\bottomrule
\end{tabular}
\captionof{table}{Summary of notation and conventions used throughout this work.}
\label{tab:notation}
\end{center}
%%%%%%%%%%%%%%%%%%%%%%%%%%%%%%%%%%%%%%%%%%%%%%%%%%%%%%%%%%%%%%%%%%%%%%%%%%%%%%%%%%%%%%%%%%
%%%%%%%%%%%%%%%%%%%%%%%%%%%%%%%%%%%%%%%%%%%%%%%%%%%%%%%%%%%%%%%%%%%%%%%%%%%%%%%%%%%%%%%%%%
%%%%%%%%%%          APPENDIX-B        %%%%%%%%%%%%%%%%%%%%%%%%%%%%%%%%%%%%%%%%%%%%%%%%%%%%
%%%%%%%%%%%%%%%%%%%%%%%%%%%%%%%%%%%%%%%%%%%%%%%%%%%%%%%%%%%%%%%%%%%%%%%%%%%%%%%%%%%%%%%%%%
%%%%%%%%%%%%%%%%%%%%%%%%%%%%%%%%%%%%%%%%%%%%%%%%%%%%%%%%%%%%%%%%%%%%%%%%%%%%%%%%%%%%%%%%%%
\setcounter{table}{0}
\renewcommand{\thetable}{B.\arabic{table}}
%\newpage
\section{Summary of boundary data and asymptotic symmetries}
\label{app:summary}
In this appendix we summarize the boundary data and the corresponding
asymptotic symmetry algebras in both the continuum and discrete
realizations of Carroll dilaton gravity. This comparison makes explicit
the correspondence between continuum boundary fields and lattice
variables, and clarifies how affine Carroll symmetries are reduced to
Virasoro-type structures under appropriate boundary conditions.
\begin{table}[t]
\centering
\begin{tabular}{lll}
\toprule
\textbf{Theory} & \textbf{Boundary data} & \textbf{ASA} \\
\midrule

Continuum Carroll BF &
$\mathcal L^I,\;\mathcal X^I$ &
Affine Carroll algebra \\

Continuum Carroll (conformal) &
$\mathcal L,\;\mathcal X$ &
Virasoro-type algebra \\

Discrete Carroll BF &
$\mathcal L_n^I,\;\mathcal X_n^I$ &
Discrete affine Carroll \\

Discrete Carroll (reduced) &
$\mathcal L_n,\;\mathcal X_n$ &
Discrete Virasoro-type \\

\bottomrule
\end{tabular}
\caption{Boundary data and asymptotic symmetry structures in
continuum and discrete Carroll dilaton gravity.}
\label{tab:summary}
\end{table}
\newpage
%%%%%%%%%%%%%%%%%%%%%%%%%%%%%%%%%%%%%%%%%%%%%%%%%%%%%%%%%%%%%%%%%%%%%%%%%%%%%%%%%%%%%%%%%%
%%%%%%%%%%%%%%%%%%%%%%%%%%%%%%%%%%%%%%%%%%%%%%%%%%%%%%%%%%%%%%%%%%%%%%%%%%%%%%%%%%%%%%%%%%
%%%%%%%%%%          APPENDIX-C        %%%%%%%%%%%%%%%%%%%%%%%%%%%%%%%%%%%%%%%%%%%%%%%%%%%%
%%%%%%%%%%%%%%%%%%%%%%%%%%%%%%%%%%%%%%%%%%%%%%%%%%%%%%%%%%%%%%%%%%%%%%%%%%%%%%%%%%%%%%%%%%
%%%%%%%%%%%%%%%%%%%%%%%%%%%%%%%%%%%%%%%%%%%%%%%%%%%%%%%%%%%%%%%%%%%%%%%%%%%%%%%%%%%%%%%%%%
\setcounter{table}{0}
\renewcommand{\thetable}{C.\arabic{table}}

\section{Discrete curvature and the flatness constraint}
\label{app:discrete_curvature}
In this appendix we derive the relation between the plaquette holonomy and 
the continuum curvature appearing in Section~\ref{sec:Discrete} 
for a general cellular decomposition of the manifold. Let $f$ be an oriented  
face of the lattice, and let its boundary $\partial f$ be the ordered closed path
\begin{equation}
\partial f=\ell_1\circ\ell_2\circ\cdots\circ\ell_n,
\end{equation}
where the number of boundary links $n$ is arbitrary.
To each oriented link $\ell$ we associate the group-valued holonomy
\begin{equation}
\mathcal U_\ell=\mathcal P\exp\!\left(\int_\ell\mathcal A\right).
\end{equation}
If the orientation of the link is reversed, one has
\begin{equation}
\mathcal U_{\ell^{-1}}=\mathcal U_\ell^{-1}.
\end{equation}

The discrete curvature associated with the face $f$ is defined
by the holonomy around its boundary,
\begin{equation}
\mathcal W_f=\prod_{\ell\in\partial f}\mathcal U_\ell
\end{equation}
with the ordered product taken according to the orientation of
$\partial f$.

%\paragraph{Small-face expansion.}
Assume that the face $f$ is sufficiently small, with typical linear
size $\varepsilon$. Choose a base point $x_0\in\partial f$ and write
the connection locally as
\begin{equation}
\mathcal A=A_\mu(x)\,dx^\mu .
\end{equation}
For each link $\ell_i$ one has the expansion
\begin{equation}
\mathcal U_{\ell_i}
=
\mathbbm{1}+\int_{\ell_i}\mathcal A+\mathcal O(\varepsilon^2),
\end{equation}
since the line integral along a single short link is of order
$\varepsilon$.

Multiplying the link holonomies around the closed path $\partial f$
and keeping terms up to order $\varepsilon^2$, one obtains
\begin{equation}
\mathcal W_f
=
\mathbbm{1}+\oint_{\partial f}\mathcal A
+\frac12\oint_{\partial f}\!\!\oint_{\partial f}\!
\mathcal P\big(\mathcal A(s_1)\mathcal A(s_2)\big)
+\mathcal O(\varepsilon^3).
\end{equation}
The first-order contribution vanishes for a closed loop up to the
variation of the connection across the face, so that the leading
nontrivial term is of order $\varepsilon^2$. Using the standard
expansion of the Wilson loop, one finds
\begin{equation}
\mathcal W_f
=
\mathbbm{1}+\int_f \mathcal F+\mathcal O(\varepsilon^3),
\end{equation}
where
\begin{equation}
\mathcal F=d\mathcal A+\mathcal A\wedge\mathcal A
\end{equation}
is the continuum curvature two-form.

Equivalently, this may be written as
\begin{equation}
\mathcal W_f
=
\exp\!\left(\int_f\mathcal F+\mathcal O(\varepsilon^3)\right).
\end{equation}
Thus, in the continuum limit, the plaquette holonomy measures the
curvature flux through the face $f$.

%\paragraph{Discrete BF action.}
In the continuum BF theory the action is
\begin{equation}
\mathcal S=\int \langle\mathcal X,\mathcal F\rangle .
\end{equation}
In the discrete theory the curvature is represented by the
Lie-algebra element $\log\mathcal W_f$. Associating a dilaton variable
$\mathcal X_f$ to each face $f$, the discrete action takes the form
\begin{equation}
\mathcal S_{\mathrm{disc}}
=
\sum_f \langle\mathcal X_f,\log\mathcal W_f\rangle +\mathcal S_{\mathrm{boundary}}.
\end{equation}

%\paragraph{Flatness constraint.}
Varying with respect to $\mathcal X_f$ gives
\begin{equation}
\delta\mathcal S_{\mathrm{disc}}
=
\sum_f \langle \delta\mathcal X_f,\log\mathcal W_f\rangle .
\end{equation}
Since $\delta\mathcal X_f$ is arbitrary, the equation of motion implies
\begin{equation}
\log\mathcal W_f=0.
\end{equation}
Choosing the branch of the logarithm continuously connected to the
identity therefore yields
\begin{equation}
\mathcal W_f=\mathbbm{1}.
\end{equation}
Hence the discrete equations of motion impose vanishing plaquette
holonomy, which is the lattice counterpart of the continuum flatness
condition
\begin{equation}
\mathcal F=0.
\end{equation}

%\paragraph{Variation with respect to the connection $\mathcal A$}

We now derive the discrete equation obtained by varying the action
with respect to the connection variables. In the lattice description
the fundamental gauge variables are the holonomies $\mathcal U_\ell$
associated with lattice links.

Let $\ell$ be a link belonging to the boundary of several faces
$f$. The variation of the plaquette holonomy $\mathcal W_f$ with respect to
$\mathcal U_\ell$ follows from the ordered product definition
\begin{equation}
\mathcal W_f=\prod_{\ell\in\partial f}\mathcal U_\ell .
\end{equation}

If the link $\ell$ appears in the boundary of $f$, the variation of
$\mathcal W_f$ is

\begin{equation}
\delta\mathcal  W_f
=
\mathcal U_{\ell_1}\cdots (\delta\mathcal  U_\ell)\cdots\mathcal  U_{\ell_n}.
\end{equation}

Using the identity
\begin{equation}
\delta(\log\mathcal  W_f)=\mathcal W_f^{-1}\delta\mathcal  W_f ,
\end{equation}
the variation of the discrete BF action becomes

\begin{equation}
\delta_{\mathcal  A}\mathcal  S_{\mathrm{disc}}
=
\sum_f
\left\langle
\mathcal X_f,
\mathcal W_f^{-1}\delta\mathcal  W_f
\right\rangle .
\end{equation}

Collecting all contributions from faces sharing the link $\ell$
one obtains the lattice equation of motion

\begin{equation}
\sum_{f\supset\ell}
\mathrm{Ad}_{\mathcal U_{\ell,f}}(\mathcal X_f)=0 ,
\end{equation}

where $\mathrm{Ad}$ denotes the adjoint action of the holonomy
transporting the dilaton from the plaquette center to the link.

%\paragraph{Continuum limit}

In the continuum limit this equation reduces to

\begin{equation}
D_{\mathcal A} X =0 ,
\end{equation}

which is the standard equation of motion of BF theory expressing
covariant constancy of the dilaton field.
%\newpage
%%%%%%%%%%%%%%%%%%%%%%%%%%%%%%%%%%%%%%%%%%%%%%%%%%%%%%%%%%%%%%%%%%%%%%%%%%%%%%%%%%%%%%%%%%
%%%%%%%%%%%%%%%%%%%%%%%%%%%%%%%%%%%%%%%%%%%%%%%%%%%%%%%%%%%%%%%%%%%%%%%%%%%%%%%%%%%%%%%%%%
%%%%%%%%%%        THE BIBLIOGRAPHY    %%%%%%%%%%%%%%%%%%%%%%%%%%%%%%%%%%%%%%%%%%%%%%%%%%%%
%%%%%%%%%%%%%%%%%%%%%%%%%%%%%%%%%%%%%%%%%%%%%%%%%%%%%%%%%%%%%%%%%%%%%%%%%%%%%%%%%%%%%%%%%%
%%%%%%%%%%%%%%%%%%%%%%%%%%%%%%%%%%%%%%%%%%%%%%%%%%%%%%%%%%%%%%%%%%%%%%%%%%%%%%%%%%%%%%%%%%

\end{document}